\documentclass[a4paper]{article}

\usepackage{INTERSPEECH2021}
\usepackage{multirow}

\title{An Attention Self-supervised Contrastive Learning based Three-stage Model for Hand Shape Feature Representation in Cued Speech}
\name{Jianrong Wang$^1$, Nan Gu$^2$, Mei Yu$^1$, Xuewei Li$^1$, Qiang Fang$^3$, Li Liu$^{4*}$\thanks{* Corresponding author}}
\address{
  $^1$College of Intelligence and Computing, Tianjin University, Tianjin, China\\
  $^2$Tianjin International Engineering Institute, Tianjin University, Tianjin, China\\
  $^3$Institute of Linguistics, Chinese Academy of Social Science, Beijing, China\\
  $^4$Shenzhen Research Institute of Big Data, the Chinese University of Hong Kong, Shenzhen, China}
\email{liuli@cuhk.edu.cn}

\begin{document}

\maketitle
\begin{abstract}

Cued Speech (CS) is a communication system for deaf people or hearing impaired people, in which a speaker uses it to aid a lipreader in phonetic level by clarifying potentially ambiguous mouth movements with hand shape and positions. Feature extraction of multi-modal CS is a key step in CS recognition. Recent supervised deep learning based methods suffer from noisy CS data annotations especially for hand shape modality. In this work, we first propose a self-supervised contrastive learning method to learn the feature representation of image without using labels. Secondly, a small amount of manually annotated CS data are used to fine-tune the first module. Thirdly, we present a module, which combines Bi-LSTM and self-attention networks to further learn sequential features with temporal and contextual information. Besides, to enlarge the volume and the diversity of the current limited CS datasets, we build a new British English dataset containing 5 native CS speakers. Evaluation results on both French and British English datasets show that our model achieves over 90\% accuracy in hand shape recognition. Significant improvements of 8.75\% (for French) and 10.09\% (for British English) are achieved in CS phoneme recognition correctness compared with the state-of-the-art.
\end{abstract}

\section{Introduction}
%is a type of visual communication system used among deaf or hearing impaired people. It 
Cued Speech (CS) \cite{1967Cued} conveys the visual form of spoken language through different hand shapes (representing consonants) and hand positions (representing vowels) near the mouth. Accurate feature extraction (including lip, hand position and hand shape) plays an important role in automatic CS recognition.

Currently, lip and hand position feature extraction has achieved good results. For example, Liu \emph{et al.} \cite{liu2017inner}, \cite{2018Visual} proposed a CLNF based inner lip feature extraction model and an ABMM based CS hand position detector obtaining a good performance. Besides, YOLOv5 \cite{glenn_jocher_2021_4679653} performs well in real-time object detection which can be used to detect hand positions in CS. However, there are still some challenges which limit a good hand shape feature extractor. Firstly, the movements of lips and hands are asynchronous in CS~\cite{2018Automatic}, \cite{liu2019novel} while most of the proposed hand shape annotation methods rely on audio based segmentation~\cite{2004Generating}. This causes noisy hand shape annotation for CS data. Secondly, since the hand shape changes quickly when it moves, for example, the same hand shape with fast moving, variable rotations, occlusion and motion blur causing different appearance, it is difficult to extract accurate and robust hand shape features \cite{rastgoo2021hand}. Indeed, most previous works~\cite{2018Visual}, \cite{2021AB}, \cite{liu2020re} ignored the variability and complexity of hand shape in CS. Thirdly, the limited CS dataset brings great limitations to the deep learning based methods of CS hand shape feature extraction. At present, the published CS datasets merely include a single speaker French dataset~\cite{2018Automatic} and a single speaker British English dataset~\cite{2019Automatic}. 
%This makes multi-lingual and multi-speaker CS recognition difficult.

To address the above mentioned problems and improve the performance of CS hand shape feature representation, in this work, we propose a three-stage hand shape feature extraction model. More precisely, in the first stage, a self-supervised contrastive learning module \cite{2020AB} is explored to learn image structure information. In the second stage, we manually annotate a small amount (10\% of the training set) of CS data to fine-tune the module, and obtain more accurate hand shape features. In the third stage, we combine Bi-LSTM \cite{2013Speech} and self-attention networks (SANs) \cite{2017Attention} to obtain temporal and contextual information of hand shapes. In addition, we establish a new British English dataset with 390 sentences recorded by five native speakers, which is the first multi-speaker British English CS dataset. We validate our model in multi-lingual and multi-speaker scenarios (on French and British English datasets), achieving better results than state-of-the-art (SOTA).

%we manually annotate a small amount (10\% of the training set) of CS data to fine-tune the self-supervised contrastive learning module
%We validate our model in multi-lingual and multi-speaker scenarios (on French and British English datasets) and find that when fine-tuned on data with only 10\% of the labels, we achieve 98\% hand shape recognition accuracy in French dataset and over 94\% in British English dataset. Besides, 79.63\% and 73.84\% accuracy are achieved in French and British English continuous CS phoneme recognition, which show significant improvements compared with previous state-of-the-art (SOTA)~\cite{2021AB}. 

In summary, our contributions are three-fold:
\begin{itemize}
\item A new multi-speaker British English CS dataset is built to solve the problem of CS hand shape recognition for the first time.
\item A three-stage hand shape feature extraction model based on self-supervised contrastive learning and self-attention mechanism is proposed to model spatial and temporal features of CS hand shape.
\item Experimental results in multi-lingual and multi-speaker scenarios show that our model is superior to SOTA in hand shape feature extraction and CS recognition.
\end{itemize}

\section{Related Work}
\begin{figure*}
  \centering
  \includegraphics[width=0.8\linewidth]{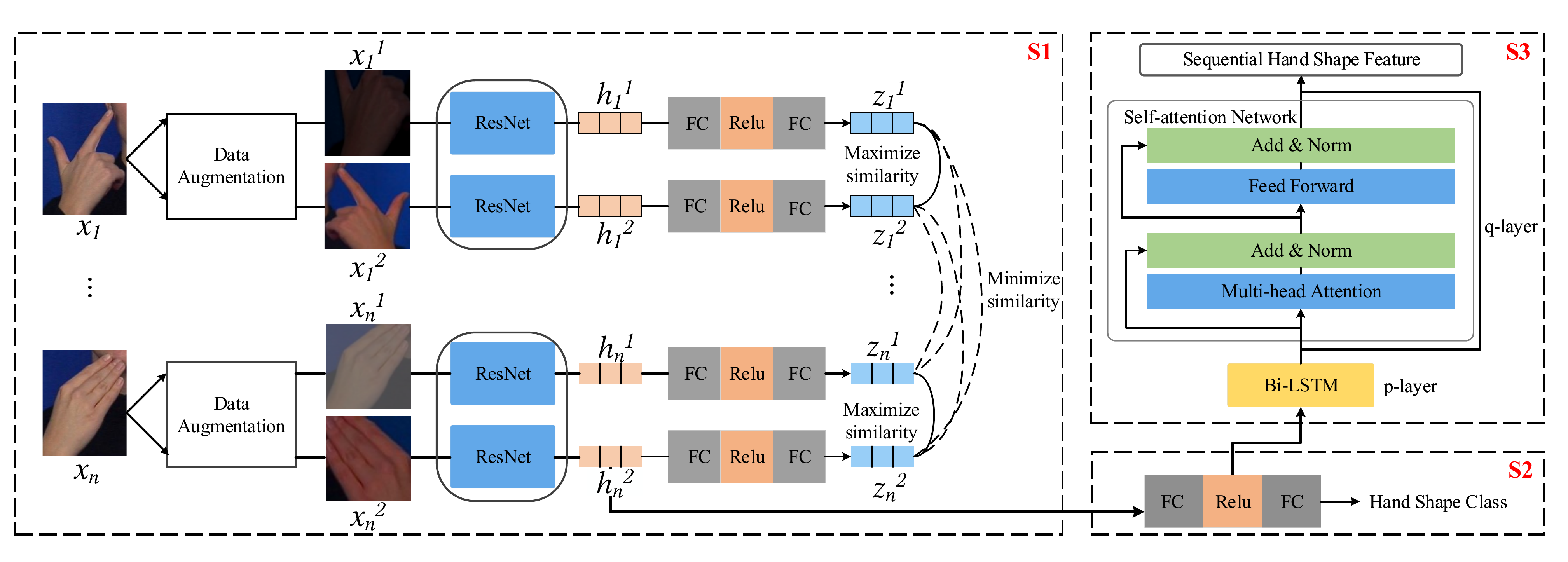}
  \caption{Overview of the proposed model. S1, S2 and S3 are three stages.}
  \label{fig:Overview}
\end{figure*}

Hand shape feature extraction plays an important role in CS recognition. In the early works, the classical methods extracted hand shape features with the artifices on the hand. Burger \emph{et al.} \cite{burger2005cued} let the speaker wear black gloves to obtain accurate hand segmentation. Noureddine \emph{et al.} \cite{2006Hand} placed blue marks on the fingers of the speaker to obtain the coordinates of fingers, so as to extract the accurate position and shape of the hand. However, both the speaker's clothing color and background color will affect the accuracy of hand segmentation.
%However, these methods prohibit the speaker from wearing clothes of the same color as the mark. Moreover, when recording video, the background can not appear the same color, otherwise it will bring great difficulties to the hand segmentation.

Recently, \cite{2018Visual} proposed a new scheme without any artifices, which used a GMM based foreground extraction for hand detection. Then, Convolutional Neural Networks (CNNs) were employed to extract features from hand regions, feeding them to an HMM-GMM classifier. Similarly, with skin color detection, \cite{2021AB} used a motion-based Kalman filter to track the hand and employed 3D-CNNs to extract spatio-temporal features from hand region sequence. However, these methods are easily affected by skin color, illumination and noisy annotation data, which can easily lead to inaccurate hand positions. 

Hand shape feature extraction is also active in other fields, such as gesture recognition \cite{mitra2007gesture} and sign language recognition \cite{rastgoo2021hand}. Current gesture recognition algorithms attempt to extract hand features such as skin color, appearance, motion, skeleton and deep learning based detection~\cite{oudah2020hand}. However, skin color based~\cite{perimal2018hand} and motion based methods~\cite{2018Visual}, \cite{pun2011real} were greatly affected by illumination changes and background color, while the hand moves around the face in CS. Appearance based methods~\cite{chen2007real}, \cite{kulkarni2010appearance} modeled visual appearance by extracting image features, and compared these parameters with the features extracted from the input image. But in CS, the same hand shape may present very different appearances, while different hand shapes may present similar appearances. In addition, it is difficult to get accurate hand key points from RGB images. Therefore, it is not suitable to apply skeleton based methods~\cite{nguyen2019neural}, \cite{konstantinidis2018sign} to CS. Deep learning based methods \cite{abavisani2019improving}, \cite{narasimhaswamy2019contextual} used supervised learning scheme, which was easily affected by noisy data.

\section{Methodology}

Our model architecture mainly includes three stages (see Figure~\ref{fig:Overview}): (i) self-supervised contrastive learning based hand shape feature representation module, (ii) a small amount of manually annotated data based fine-tuning module, and (iii) self-attention based sequential hand shape feature learning module. 
% We describe each module in detail in this section.

\subsection{Self-supervised contrastive learning based hand shape feature representation}

Firstly, we perform a random transformation on input images, including cropping, horizontal flipping, color distortions, gray level distortions, and Gaussian blur. Then, we get two related images at the same instance, denoted by \(x_i^j(j=1,2)\). Next, ResNet18 is used as the basic encoder to extract the representation vectors \(h_i^j\) from the augmented data, where
\begin{equation}
  h_i^j = f(x_i^j) = ResNet(x_i^j).
  \label{eq1}
\end{equation}

Then, feature projection is used to improve the quality of the feature representation \(h_i^j\), thus mapping the feature representation to the space where contrastive loss can be applied. Two non-linear fully connected (FC) layers are used for mapping to obtain new feature representations \(z_i^j\) which is defined as 
\begin{equation}
  z_i^j = g(h_i^j) = W^{(2)}\sigma(W^{(1)}h_i^j),
  \label{eq2}
\end{equation}
where \(\sigma\) is the ReLU activation function. \(W^{(1)}\) and \(W^{(2)}\) are learnable weight matrices. Then, given a set of samples \(\{x_k\}\) and an image \(x_i\), the goal of the contrastive learning is to detect the most similar image \(x_j\) from \(\{x_k\}_{k\neq i}\).

We input \(N\) samples into this module and get \(2N\) data representations. For a positive pair, the other \(2(N-1)\) augmented examples within a batch are regarded as negative examples. The cosine similarity \(sim(\cdot)\) is used to calculate the probability of similarity between any two feature representations. After that, a contrastive loss function is applied to calculate the loss of positive pairs, which is defined as
\begin{equation}
  \ell _{i, j} = -\log{\frac{\exp (sim(z_i, z_j)/t)}{\sum_{k=1}^{2N}1_{[k\neq i]}\exp (sim(z_i, z_k)/t)}},
  \label{eq3}
\end{equation}
where \(1_{[k\neq i]}\) is an indicator function and its value is 1 iff \(k\neq i\). \(t\) represents the temperature parameter. We calculate the loss of the sum of all positive pairs in a batch and take the average.

\subsection{Manually annotated data based fine-tuning}

We fine-tune the model with a small amount of manually annotated data (10\% of the training set) to obtain more accurate hand shape features. In the feature projection, to map feature representations to the same space, some information of hand shapes are discarded. Therefore, only the basic encoder network in the first module is used for fine-tuning. We directly feed hand ROI to the pre-trained ResNet18, obtaining the hand shape features and feeding them to this module. Two non-linear FC layers are used as the classification network in this module. 

\subsection{Self-attention based sequential hand shape feature learning}
Based on the above two modules, we can extract accurate spatial hand shape features. However, CS hand shape is a video sequence, thus contextual information is also important. To learn temporal and sequential information of hand shapes, we combine Bi-LSTM and multi-head SANs. Among them, Bi-LSTM models the global information of a sequence to extract features with temporal and contextual information. SANs learn the internal dependence of the sequence, so as to better model the internal structure of the sequence and emphasize the important information. In order to make full use of the features learned from Bi-LSTM and SANs, we use a short-cut connection to add their outputs. Let \(C=\{c_1, c_2, ..., c_N\}\) be the input sequence, the representation of the encoder is defined as:
\begin{equation}
  s_{output} = s^p + s^q,
  \label{eq4}
\end{equation}
\begin{equation}
  s^p = U(C), \text{ and } s^q = V(s^p),
  \label{eq5}
\end{equation}
where \(p\) and \(q\) represent the layer number of the encoder, respectively. \(U\) and \(V\) represent the Bi-LSTM and SANs, respectively. \(s_{output}\) represents the obtained sequential hand shape features.

% \begin{figure}[t]
%   \centering
%   \includegraphics[width=0.8\linewidth]{./hand_imgs/seq_fea.pdf}
%   \caption{The architecture of the Sentence-level feature extractor. The yellow part indicates the Bi-LSTM and the blue part indicates the SANs. The purple part is the non-linear fully connected layer used for phoneme classification.}
%   \label{fig:BILSTM}
% \end{figure}

\section{Evaluations}

% In this section, we discuss our datasets, implementation details, and experimental results.

\subsection{Dataset}

Two public single-speaker French (named LM) and British English (named CA) CS datasets contain 238 sentences (repeated twice) and 97 sentences, respectively. The French dataset contains 8 visemes, 8 hand shapes and 5 hand positions to encode 14 vowels and 20 consonants. More details can be referred to \cite{2018Automatic} and \cite{2019Automatic}. 

Based on CA, which was published by the author in 2019, we build a new five-speaker British English dataset (the original data contain video and audio). With forced alignment, the acoustic signal synchronized with the video is automatically marked. Then, we post-check them manually. The expanded datasets were recorded by 4 different professional CS interpreters (named EM, KA, LD and VK, respectively). The specific data is shown in Table~\ref{tab:corpora}. The new British English dataset contains 11 visemes \cite{jachimski2018comparative}, 8 hand shapes and 4 hand positions to encode 17 vowels and 24 consonants. RGB video images of the interpreter's upper body are available at 25 fps, and the spatial resolution is 720 × 1280.

% The public sing-speaker French CS dataset \cite{2018Automatic} includes 238 sentences (repeated twice) recorded by a professional CS interpreter.RGB video images of the interpreter's upper body are provided at 50 fps, and the spatial resolution is 720 × 576. It contains 8 lip shapes, 8 hand shapes and 5 hand positions, to encode 14 vowels and 20 consonants. It should be noted that the British English CS dataset is different in that it contains 11 lip shapes \cite{jachimski2018comparative}, 8 hand shapes and 4 hand positions, and encodes 17 vowels and 24 consonants. RGB video images of the interpreter's upper body are available at 25 fps, and the spatial resolution is 720 × 1280.

% In \cite{2019Automatic}, a British English dataset that contains only 97 sentences and recorded by only one person (named CA) was published. With that, we build a brand new British English dataset (the original datas are only video and audio). With forced alignment, the acoustic signal synchronized with the video is automatically marked. Then, we post-check them manually. The expanded datasets were recorded by 4 different professional CS interpreters (named EM, KA, LD and VK respectively). The specific data is shown in Table~\ref{tab:corpora}. 

\subsection{Implementation Details}

In the following experiments, we randomly select 80\% of the dataset as the training set and 20\% of the dataset as the test set. Five groups of data are randomly generated for cross validation.

\subsubsection{Hand shape feature extraction}

We use our model to extract hand shape features from hand ROI. In stage 1, all training set (with noisy labels) and the Adam optimizer are used to train the model for 200 epochs (the English model is trained on the French pre-trained model for 100 epochs). In stage 2, a small amount of manually annotated data, the Adam optimizer and cross-entropy loss function are used to train 60 epochs with a learning rate of 1e-4. Then, we feed the continuous hand ROI into our model and feed the features obtained from stage 2 into stage 3. In stage 3, Bi-LSTM (2 layers) and multi-head SANs (3 layers and 16 heads) are used to extract the continuous hand shape features. The Connectionist Temporal Classification (CTC) \cite{graves2006connectionist} is used as the loss function. Empirically, the learning rate is 0.001 and the batch size is 1.

\begin{table}[t]
  \caption{A summary of the French and British English datasets.}
  \label{tab:corpora}
  \centering
  \setlength{\tabcolsep}{0.7mm}{
  \begin{tabular}{ccc}
    \toprule
    % \textbf{Speaker}      & \textbf{Speech amount(sentence)}       &\textbf{Language}\\
%     \multirow{2}{*}{Speaker} & Speech amount & \multirow{2}{*}{Language} \\
%   {} &  (sentence) &  {} \\
    \textbf{Speaker} & \textbf{Speech amount} & \textbf{Language} \\
    {} &  \textbf{(sentence)} &  {} \\
    \midrule
    LM                       & 238x2             & French                                      \\
    CA                     & 97                & British English                                 \\
    EM                     & 100               & British English                                 \\
    KA                    & 100               & British English                           \\
    LD                  & 50                & British English                            \\
    VK                    & 43                & British English                            \\
    \bottomrule
  \end{tabular}}
\end{table}

\subsubsection{Continuous CS phoneme recognition}

We detect the position of lips and hands at first. The facial landmark detector in the open source library dlib \cite{2009Dlib} is used to extract the key points of the outer edge of lips. Then, the lip ROI is obtained based on these key points. YOLOv5 is used to detect hand positions. We pre-train the model on a public hand dataset \cite{mittal2011hand}. Next, we mark a small number of hand bounding boxes on CS datasets to fine-tune the model. Finally, the hand positions and ROI can be  obtained simultaneously. After that, we extract the features of the three streams, respectively. We employ a 2D-CNN which contains two convolutional layers (8 filters, a kernel size of 7×7 pixels, a down-sampling factor of 3) to extract lip features from lip ROI, and also employ an ANN to extract hand position features from coordinate values. The method of obtaining static hand shape features is described in 4.2.1. We concatenate the features of the three streams. Then, the S3 module which has the same structure and setting as the continuous hand shape feature extraction experiment is used to recognize continuous CS phoneme.
% We obtain feature vectors of 64, 128, and 64 dimensions from the three feature streams and concatenate them to get a 256-dimensional feature vector. Then, the stage 3 module which has the same structure and setting as the continuous hand shape feature extraction experiment is used to recognize continuous CS phoneme.

\subsubsection{Evaluation metrics}

In the hand shape recognition based on a single image, the classification accuracy (\(acc = \frac{n_{correct}}{n_{total}}\)) is used as the evaluation metrics, where \(n_{correct}\) and \(n_{total}\) represent the number of correct samples and the total number of samples, respectively. In continuous hand shape recognition and CS phoneme recognition, phone error rate (\(T_e = \frac{n_i + n_d + n_s}{N}\)), and phone correct rate (\(T_c=1-T_e\)) are used as the evaluation metrics, where \(n_i\), \(n_d\) and \(n_s\) represent the number of insert, delete and replace errors, respectively. \(N\) represents the input sequence length.
% More precisely, they can be defined as follows:
%calculate the Levenshtein distance between the prediction sequence and the reference sequence
% % 
% \begin{equation}
%   acc = \frac{n_{correct}}{n_{total}},
%   \label{eq7}
% \end{equation}
% %
% % 
% \begin{equation}
%   T_e = \frac{n_i + n_d + n_s}{N},
%   \label{eq8}
% \end{equation}
% %
% % 
% \begin{equation}
%   T_c = 1-T_e,
%   \label{eq9}
% \end{equation}
% %

\subsection{Result and Discussion}

We report the results of CS hand shape recognition and phoneme recognition quantitatively and qualitatively.

\subsubsection{Hand shape recognition}

To verify the influence of noisy data on hand shape recognition performance, we compare ResNet18 directly trained with noisy data (100\% of the training set) with ResNet18 trained with manually annotated data (10\% of the training set). As shown in Table~\ref{tab:hand}, ResNet18 trained with manually annotated data achieves over 30\% improvement on both French and British English CS datasets. Then, we compare our model which only contains S1 and S2 with ResNet18 trained with manually annotated data. In both our model and ResNet18, we use the same amount of manually annotated data. It can be seen from Table~\ref{tab:hand} that our model performs better because the self-supervised contrastive learning module makes full use of noisy data for feature representation. Finally, in order to verify the performance of our model in multi-lingual and multi-speaker scenarios, we mix French and British English CS datasets for training. We take out 80\% of each dataset as the mixed training set. Obviously, better results are achieved due to the increase in the size and diversity of the dataset. The results show that our model is not affected by multiple speakers. We notice that there is a slight drop in EM's result, which may be caused by the small amount of EM's data.

\begin{table}[t]
  \caption{Comparison results (acc\%) using ResNet18 based on supervised learning, and our model. We use noisy data (n) and manually annotated data (a) to train ResNet18. Ours-S3 represents our model without S3 and Ours-S3(multi) represents our model in multi-lingual and multi-speaker scenarios.}
  \label{tab:hand}
  \centering
  \setlength{\tabcolsep}{0.8mm}{
  \begin{tabular}{ccccccc}
    \toprule
    \textbf{Method}      & \textbf{LM}      & \textbf{CA}       & \textbf{EM}      & \textbf{KA}      & \textbf{LD}        & \textbf{VK}\\
    \midrule                                                
    % 2D-CNN                & 62.85            & 56.77             & 58.04              & 63.24               & 51.25                   & 51.93  \\
    ResNet18(n)         & 61.33            & 53.92             & 54.58              & 60.14               & 47.68                   & 48.86  \\
    ResNet18(a)         & 94.32            & 89.17             & 83.58              & 91.49               & 86.61                   & 86.55   \\
    Ours-S3              & 96.39            & 95.05             & \textbf{91.59}     & 96.45               & 92.56                   & 95.18  \\
    Ours-S3(multi)                  & \textbf{96.50}   & \textbf{95.73}    & 91.54              & \textbf{96.52}      & \textbf{92.89}          & \textbf{96.28}  \\
    \bottomrule
  \end{tabular}}
\end{table}

\begin{figure}[t]
  \centering
  \includegraphics[width=0.85\linewidth]{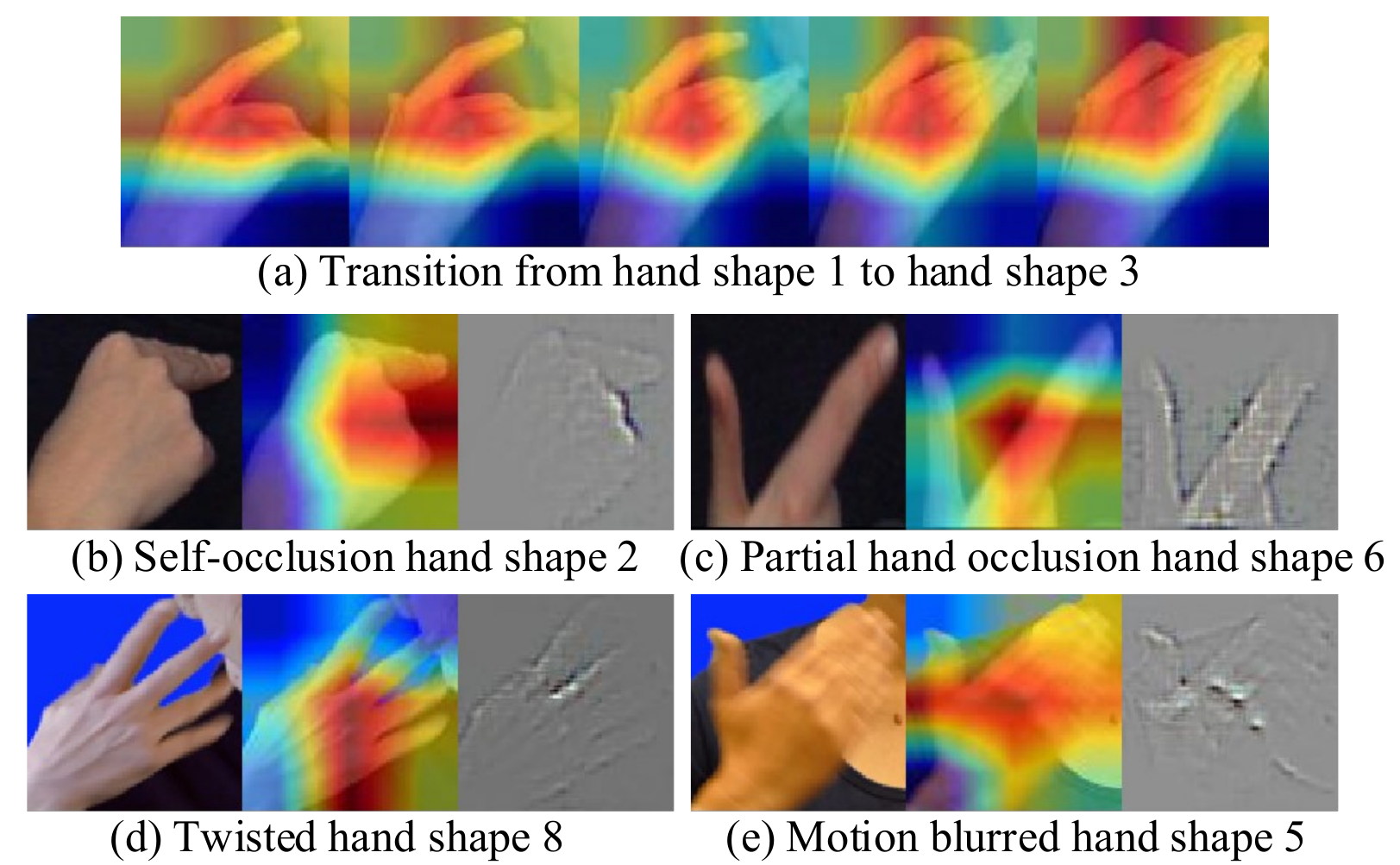}
  \caption{Feature visualization of hand shapes, where (a) is the Grad-CAM++ \cite{chattopadhay2018grad} visual graph. The images of each group in (b)$\sim$(e) are original image, Grad-CAM \cite{selvaraju2017grad} visual graph and Guided Grad-CAM \cite{selvaraju2017grad} visual graph, respectively.}
  \label{fig:hand_trans}
\end{figure}

In Figure~\ref{fig:hand_trans}, to verify the effectiveness of the extracted hand shape features, we qualitatively visualize the feature map of hands in complex scenes. It can be seen from (a) that the hand shape is distorted during transition, but our model still captures the key regions. In (b), the index finger and the middle finger are overlapped, but the features of the index and middle finger are both well learned. At the beginning and end of a sentence, the hand is gradually raised or put down. Therefore, some hands are only partially exposed in this process, as shown in (c). However, even (c) only shows two fingers, the hand shape can be correctly recognized. (d) is a twisted hand shape 8. It can be seen that even though the hand shape is similar to hand shape 4, the features are still well learned. And (e) is a hand shape 5 blurred by motion. Although the hand shape is very fuzzy, our model can still reconstruct the hand shape.

The recognition results of continuous hand shape in each dataset can be seen in Table~\ref{tab:contious_hand}. Each time the hand shape sequence in a sentence is identified. Compared with SOTA (i.e., CNN-HMM) \cite{2018Visual}, a significant improvement of 16.42\% correctness is achieved in the French dataset. In the British English dataset, our model also achieves high accuracy. Ablation study (Ours-S2 and Ours-SANs) shows that all modules in our model are important. In addition, more accurate results can be obtained with manually annotated data. Compared with the model trained with noisy data (i.e., ReNet18(n)+S3), our model achieves much higher hand shape recognition correctness.
% Notably, because continuous hand shape recognition is based on sentences, the limited size of datasets limits the recognition performance.
\begin{table}[t]
  \caption{Comparison of continuous hand shape recognition results in \(T_c\) (\%). CNN-HMM \cite{2018Visual} is the SOTA in French dataset. ResNet18(n)+S3 represents the model which combines ResNet18(n) and S3. Ours-S2  represents our model without S2. Ours-SANs represents our model without SANs. Ours represents our model.}
  \label{tab:contious_hand}
  \centering
  \setlength{\tabcolsep}{0.8mm}{
  \begin{tabular}{ccccccc}
    \toprule
    \textbf{Method}      & \textbf{LM}      & \textbf{CA}       & \textbf{EM}      & \textbf{KA}      & \textbf{LD}        & \textbf{VK}\\
    \midrule   
    CNN-HMM \cite{2018Visual} & 68.50       & ---                 & ---               & ---          & ---            & ---  \\
    ResNet18(n)+S3               & 83.99        & 78.16          & 76.15         & 76.08         & 74.42        & 67.52\\
    Ours-S2            & 84.81         & 82.27       & 79.13            & 83.57                   & 75.86        & 75.44\\
    Ours-SANs             & 84.88      & 84.75        & 83.81       & 89.40       & 80.47       & 76.12       \\
    Ours        & \textbf{84.92}      & \textbf{85.25}        & \textbf{85.19}       & \textbf{89.52}       & \textbf{82.22}       & \textbf{78.94}     \\
    \bottomrule
  \end{tabular}}
\end{table}

\subsubsection{Phoneme recognition}

Finally, we consider CS phoneme recognition, which uses the lip, hand position, and hand shape as the input, and the output is the phoneme class (33  and 41 phonemes in French and British English, respectively). The comparison under different models (see Table~\ref{tab:phoneme}) shows that our model achieves the best results on both French and British English datasets. Compared with SOTA \cite{2021AB}, 8.75\% and 10.09\% improvements are achieved in the two datasets, respectively. This verifies that our model can still perform well in the multi-lingual scenario. There are significant differences in performance for each dataset. This may be due to the limited size of LD and VK datasets. But the exciting finding is that the model with SANs outperforms other models in Table~\ref{tab:phoneme}. The above results show that our model can reduce the influence of noisy data on CS phoneme recognition.

To verify the performance of our model in the multi-speaker scenario, we mix data from five speakers in the British English dataset. 80\% data of each speaker is used as the training set. It can be observed that all the results are greatly improved because the increasing diversity of dataset is beneficial to self-supervised contrastive learning.

\begin{table}[h]
  \caption{Comparison of phoneme recognition results in \(T_e\) (\%). 3D-CNN+TDS \cite{2021AB} is the SOTA. Ours(multi) represents our model in the multi-speaker scenario.}
  \label{tab:phoneme}
  \centering
  \setlength{\tabcolsep}{0.75mm}{
  \begin{tabular}{ccccccc}
    \toprule
    \textbf{Method}      & \textbf{LM}      & \textbf{CA}       & \textbf{EM}      & \textbf{KA}      & \textbf{LD}        & \textbf{VK}\\
    \midrule   
    3D-CNN+TDS \cite{2021AB}  & 29.12             & 36.25        & ---           & ---                  & ---                & ---       \\
    ResNet18(n)+S3         & 23.27             & 30.48          & 27.36            & 26.20              & 46.79            & 46.60     \\
    Ours-SANs               & 21.8              & 27.21          & 25.41            &25.96               & 43.89            & 45.42            \\
    Ours                    & \textbf{20.37}    & 26.16          & 23.39            & 24.88              & 38.42            & 42.02    \\
    Ours(multi)            & ---                & \textbf{19.93}       & \textbf{17.64}      & \textbf{14.59}    & \textbf{20.77}    & \textbf{26.38} \\
    \bottomrule
  \end{tabular}}
\end{table}

% \begin{table}[h!]
%   \caption{The phoneme recognition results in \(T_e\) (\%) of 5 people's mixed training in British English dataset. }
%   \label{tab:mix_phoneme}
%   \centering
%   \setlength{\tabcolsep}{0.8mm}{
%   \begin{tabular}{ccccccc}
%     \toprule
%     \textbf{Method}    & \textbf{All}   & \textbf{CA}       & \textbf{EM}      & \textbf{KA}      & \textbf{LD}        & \textbf{VK}\\
%     \midrule   
%     % Ours-SANs             & 20.47          & 22.79             & 17.67               & 16.74               & 20.75                   & 30.16            \\
%     Ours        & 19.38          & 19.93             & 17.64               & 14.59               & 20.77                   & 26.38                                            \\
%     \bottomrule
%   \end{tabular}}
% \end{table}

\section{Conclusions}

In this work, we propose an attention self-supervised contrastive learning based three-stage hand shape feature extraction model. Our model solves the problem caused by excessive noisy annotation in CS hand shape feature extraction. Experimental results show superior performance to SOTA methods. As for future work, we will explore the effective sample pair selection methods for self-supervised contrastive learning in CS video.

\section{Acknowledgements}

This work was supported by the GuangDong Basic and Applied Basic Research Foundation (No. 2020A1515110376) and the National Natural Science Foundation of China (grant No. 61977049). The authors would like to thank the professional CS speakers from Cued Speech UK for the British English CS dataset recording.
% The ISCA Board would like to thank the organizing committees of the past INTERSPEECH conferences for their help and for kindly providing the template files. \\
% Note to authors: Authors should not use logos in the acknowledgement section; rather authors should acknowledge corporations by naming them only.

\bibliographystyle{IEEEtran}

\bibliography{mybib}

\begin{thebibliography}{10}
\providecommand{\url}[1]{#1}
\csname url@samestyle\endcsname
\providecommand{\newblock}{\relax}
\providecommand{\bibinfo}[2]{#2}
\providecommand{\BIBentrySTDinterwordspacing}{\spaceskip=0pt\relax}
\providecommand{\BIBentryALTinterwordstretchfactor}{4}
\providecommand{\BIBentryALTinterwordspacing}{\spaceskip=\fontdimen2\font plus
\BIBentryALTinterwordstretchfactor\fontdimen3\font minus
  \fontdimen4\font\relax}
\providecommand{\BIBforeignlanguage}[2]{{%
\expandafter\ifx\csname l@#1\endcsname\relax
\typeout{** WARNING: IEEEtran.bst: No hyphenation pattern has been}%
\typeout{** loaded for the language `#1'. Using the pattern for}%
\typeout{** the default language instead.}%
\else
\language=\csname l@#1\endcsname
\fi
#2}}
\providecommand{\BIBdecl}{\relax}
\BIBdecl

\bibitem{1967Cued}
R.~O. Cornett, ``Cued speech,'' \emph{American annals of the deaf}, vol. 112,
  no.~1, pp. 3--13, 1967.

\bibitem{liu2017inner}
L.~Liu, G.~Feng, and D.~Beautemps, ``Inner lips feature extraction based on
  clnf with hybrid dynamic template for cued speech,'' \emph{EURASIP Journal on
  Image and Video Processing}, vol. 2017, no.~1, pp. 1--15, 2017.

\bibitem{2018Visual}
L.~Liu, T.~Hueber, G.~Feng, and D.~Beautemps, ``Visual recognition of
  continuous cued speech using a tandem cnn-hmm approach.'' \emph{in Proc.
  Interspeech}, pp. 2643--2647, 2018.

\bibitem{glenn_jocher_2021_4679653}
\BIBentryALTinterwordspacing
G.~Jocher, A.~Stoken, J.~Borovec, NanoCode012, A.~Chaurasia, TaoXie,
  L.~Changyu, A.~V, Laughing, tkianai, yxNONG, A.~Hogan, lorenzomammana,
  AlexWang1900, J.~Hajek, L.~Diaconu, Marc, Y.~Kwon, oleg, wanghaoyang0106,
  Y.~Defretin, A.~Lohia, ml5ah, B.~Milanko, B.~Fineran, D.~Khromov, D.~Yiwei,
  Doug, Durgesh, and F.~Ingham, ``{ultralytics/yolov5: v5.0 - YOLOv5-P6 1280
  models, AWS, Supervise.ly and YouTube integrations},'' Apr. 2021. [Online].
  Available: \url{https://doi.org/10.5281/zenodo.4679653}
\BIBentrySTDinterwordspacing

\bibitem{2018Automatic}
L.~Liu, G.~Feng, and D.~Beautemps, ``Automatic temporal segmentation of hand
  movements for hand positions recognition in french cued speech,'' \emph{in
  Proc. ICASSP}, pp. 3061--3065, 2018.

\bibitem{liu2019novel}
L.~Liu, G.~Feng, D.~Beautemps, and X.-P. Zhang, ``A novel resynchronization
  procedure for hand-lips fusion applied to continuous french cued speech
  recognition,'' \emph{in Proc. EUSIPCO}, pp. 1--5, 2019.

\bibitem{2004Generating}
S.~Tranter, K.~Yu, G.~Everinann, and P.~C. Woodland, ``Generating and
  evaluating segmentations for automatic speech recognition of conversational
  telephone speech,'' \emph{in Proc. ICASSP}, pp. 753--756, 2004.

\bibitem{rastgoo2021hand}
R.~Rastgoo, K.~Kiani, and S.~Escalera, ``Hand pose aware multimodal isolated
  sign language recognition,'' \emph{Multimedia Tools and Applications},
  vol.~80, no.~1, pp. 127--163, 2021.

\bibitem{2021AB}
K.~Papadimitriou and G.~Potamianos, ``A fully convolutional sequence learning
  approach for cued speech recognition from videos,'' \emph{in Proc. EUSIPCO},
  pp. 326--330, 2021.

\bibitem{liu2020re}
L.~Liu, G.~Feng, D.~Beautemps, and X.-P. Zhang, ``Re-synchronization using the
  hand preceding model for multi-modal fusion in automatic continuous cued
  speech recognition,'' \emph{IEEE Transactions on Multimedia}, vol.~23, pp.
  292--305, 2020.

\bibitem{2019Automatic}
L.~Liu, J.~Li, G.~Feng, and X.~S. Zhang, ``Automatic detection of the temporal
  segmentation of hand movements in british english cued speech.'' \emph{in
  Proc. Interspeech}, pp. 2285--2289, 2019.

\bibitem{2020AB}
T.~Chen, S.~Kornblith, M.~Norouzi, and G.~Hinton, ``A simple framework for
  contrastive learning of visual representations,'' \emph{in Proc. ICML}, pp.
  1597--1607, 2020.

\bibitem{2013Speech}
A.~Graves, A.~Mohamed, and G.~E. Hinton, ``Speech recognition with deep
  recurrent neural networks,'' \emph{in Proc. ICASSP}, pp. 6645--6649, 2013.

\bibitem{2017Attention}
A.~Vaswani, N.~Shazeer, N.~Parmar, J.~Uszkoreit, L.~Jones, A.~N. Gomez,
  L.~Kaiser, and I.~Polosukhin, ``Attention is all you need,'' \emph{arXiv
  preprint arXiv:1706.03762}, 2017.

\bibitem{burger2005cued}
T.~Burger, A.~Caplier, and S.~Mancini, ``Cued speech hand gestures recognition
  tool,'' \emph{in Proc. EUSIPCO}, pp. 1--4, 2005.

\bibitem{2006Hand}
N.~Aboutabit, D.~Beautemps, and L.~Besacier, ``Hand and lip desynchronization
  analysis in french cued speech: Automatic temporal segmentation of hand
  flow,'' \emph{in Proc. ICASSP}, pp. 633--636, 2006.

\bibitem{mitra2007gesture}
S.~Mitra and T.~Acharya, ``Gesture recognition: A survey,'' \emph{IEEE
  Transactions on Systems, Man, and Cybernetics, Part C (Applications and
  Reviews)}, vol.~37, no.~3, pp. 311--324, 2007.

\bibitem{oudah2020hand}
M.~Oudah, A.~Al-Naji, and J.~Chahl, ``Hand gesture recognition based on
  computer vision: a review of techniques,'' \emph{Journal of Imaging}, vol.~6,
  no.~8, p.~73, 2020.

\bibitem{perimal2018hand}
M.~Perimal, S.~Basah, M.~Safar, and H.~Yazid, ``Hand-gesture
  recognition-algorithm based on finger counting,'' \emph{Journal of
  Telecommunication, Electronic and Computer Engineering}, vol.~10, no. 1-13,
  pp. 19--24, 2018.

\bibitem{pun2011real}
C.-M. Pun, H.-M. Zhu, and W.~Feng, ``Real-time hand gesture recognition using
  motion tracking,'' \emph{International Journal of Computational Intelligence
  Systems}, vol.~4, no.~2, pp. 277--286, 2011.

\bibitem{chen2007real}
Q.~Chen, N.~D. Georganas, and E.~M. Petriu, ``Real-time vision-based hand
  gesture recognition using haar-like features,'' \emph{in Proc. IMTC}, pp.
  1--6, 2007.

\bibitem{kulkarni2010appearance}
V.~S. Kulkarni and S.~Lokhande, ``Appearance based recognition of american sign
  language using gesture segmentation,'' \emph{International Journal on
  Computer Science and Engineering}, vol.~2, no.~03, pp. 560--565, 2010.

\bibitem{nguyen2019neural}
X.~S. Nguyen, L.~Brun, O.~L{\'e}zoray, and S.~Bougleux, ``A neural network
  based on spd manifold learning for skeleton-based hand gesture recognition,''
  \emph{in Proc. CVPR}, pp. 12\,036--12\,045, 2019.

\bibitem{konstantinidis2018sign}
D.~Konstantinidis, K.~Dimitropoulos, and P.~Daras, ``Sign language recognition
  based on hand and body skeletal data,'' \emph{in Proc. 3DTV-CON}, pp. 1--4,
  2018.

\bibitem{abavisani2019improving}
M.~Abavisani, H.~R.~V. Joze, and V.~M. Patel, ``Improving the performance of
  unimodal dynamic hand-gesture recognition with multimodal training,''
  \emph{in Proc. CVPR}, pp. 1165--1174, 2019.

\bibitem{narasimhaswamy2019contextual}
S.~Narasimhaswamy, Z.~Wei, Y.~Wang, J.~Zhang, and M.~Hoai, ``Contextual
  attention for hand detection in the wild,'' \emph{in Proc. CVPR}, pp.
  9567--9576, 2019.

\bibitem{jachimski2018comparative}
D.~Jachimski, A.~Czyzewski, and T.~Ciszewski, ``A comparative study of english
  viseme recognition methods and algorithms,'' \emph{Multimedia Tools and
  Applications}, vol.~77, no.~13, pp. 16\,495--16\,532, 2018.

\bibitem{graves2006connectionist}
A.~Graves, S.~Fern{\'a}ndez, F.~Gomez, and J.~Schmidhuber, ``Connectionist
  temporal classification: labelling unsegmented sequence data with recurrent
  neural networks,'' \emph{in Proc. ICML}, pp. 369--376, 2006.

\bibitem{2009Dlib}
D.~E. King, ``Dlib-ml: A machine learning toolkit,'' \emph{The Journal of
  Machine Learning Research}, vol.~10, pp. 1755--1758, 2009.

\bibitem{mittal2011hand}
A.~Mittal, A.~Zisserman, and P.~H. Torr, ``Hand detection using multiple
  proposals.'' \emph{in Proc. BMVC}, pp. 1--11, 2011.

\bibitem{chattopadhay2018grad}
A.~Chattopadhay, A.~Sarkar, P.~Howlader, and V.~N. Balasubramanian,
  ``Grad-cam++: Generalized gradient-based visual explanations for deep
  convolutional networks,'' \emph{in Proc. WACV}, pp. 839--847, 2018.

\bibitem{selvaraju2017grad}
R.~R. Selvaraju, M.~Cogswell, A.~Das, R.~Vedantam, D.~Parikh, and D.~Batra,
  ``Grad-cam: Visual explanations from deep networks via gradient-based
  localization,'' \emph{in Proc. ICCV}, pp. 618--626, 2017.

\end{thebibliography}

% \begin{thebibliography}{9}
% \bibitem[1]{Davis80-COP}
%   S.\ B.\ Davis and P.\ Mermelstein,
%   ``Comparison of parametric representation for monosyllabic word recognition in continuously spoken sentences,''
%   \textit{IEEE Transactions on Acoustics, Speech and Signal Processing}, vol.~28, no.~4, pp.~357--366, 1980.
% \bibitem[2]{Rabiner89-ATO}
%   L.\ R.\ Rabiner,
%   ``A tutorial on hidden Markov models and selected applications in speech recognition,''
%   \textit{Proceedings of the IEEE}, vol.~77, no.~2, pp.~257-286, 1989.
% \bibitem[3]{Hastie09-TEO}
%   T.\ Hastie, R.\ Tibshirani, and J.\ Friedman,
%   \textit{The Elements of Statistical Learning -- Data Mining, Inference, and Prediction}.
%   New York: Springer, 2009.
% \bibitem[4]{YourName17-XXX}
%   F.\ Lastname1, F.\ Lastname2, and F.\ Lastname3,
%   ``Title of your INTERSPEECH 2021 publication,''
%   in \textit{Interspeech 2021 -- 20\textsuperscript{th} Annual Conference of the International Speech Communication Association, September 15-19, Graz, Austria, Proceedings, Proceedings}, 2020, pp.~100--104.
% \end{thebibliography}

\end{document}